# Terrestrial planet formation from a ring: long-term simulations accounting for the giant planet instability

J.M.Y. Woo[a,b], D. Nesvorný[c], J. Scora[d,e], A. Morbidelli[f,a]

[a]Laboratoire Lagrange, Université Cote d'Azur, CNRS, Observatoire de la Côte d'Azur, Boulevard de l'Observatoire, 06304 Nice Cedex 4, France
[b]Institut für Planetologie, University of Münster, Wilhelm-Klemm-Straße 10, 48149, Germany
[c]Department of Space Studies, Southwest Research Institute, 1050 Walnut St., Suite 300, Boulder, CO 80302, USA
[d] David A. Dunlap Department of Astronomy and Astrophysics, University of Toronto
[e] Sidrat Research, Toronto
[f] Collège de France, CNRS, PSL Univ., Sorbonne Univ., Paris, 75014, France

**Abstract**
The process leading to the formation of the terrestrial planet remains elusive. In a previous publication, we have shown that, if the first generation of planetesimals forms in a ring at ~1 AU and the gas disk's density peaks at the ring location, planetary embryos of a few martian masses can grow and remain in the ring. In this work, we extend our simulations beyond the gas-disk stage, covering ~200 Myr and accounting for the phase of giant planet instability, assumed to happen at different times. About half of the simulations form a pair of Venus and Earth analogues and, independently, about 10% form a Mars analogue. We find that the timing of the giant planet instability affects statistically the terrestrial system's excitation state and the timing of the last giant impacts. Hence a late instability (~60 to 100 Myr after the Solar system's birth) is more consistent with a late Moon-formation time, as suggested by radioactive chronometers. However, the late veneer mass (LVM: mass accreted after the last giant impact) of Earth-sized planets suffering a giant impact after 80 My is usually an order of magnitude lower than the value inferred from geochemistry. In addition, the final angular momentum deficit (AMD) of the terrestrial planets tends to be too high. We tested the effect on the final AMD of the generation of debris during collisions and found that it is too small to change these conclusions. We argue that the best-case scenario is that the Moon-forming event occurred between 50 and 80 My, possibly just following the giant planet instability.

1. **Introduction**

In the last 15 years it has become progressively clear that the architecture of the terrestrial planet system, with small Mercury and Mars at the extremes and massive Earth and Venus in the middle, requires that the original mass distribution was concentrated in a ring at ~1 AU (Hansen, 2009; Morishima et al., 2008; Walsh and Levison, 2016; Lykawka, 2020, Nesvorný et al., 2021; Izidoro et al., 2022). Such a concentration of mass is in striking contrast with the classical minimum mass solar nebula model, which assumes a smooth radial profile of the surface density of the disk. Some works have tried to solve this contradiction by invoking the effects of the migration of Jupiter (Walsh et al., 2011), of an early and violent giant planet instability (Clement et al., 2018), or of the chaotic motion of Jupiter and Saturn in/near the 2/1 resonance (Izidoro et al., 2016; Lykawka and Ito, 2023), all of which can deplete the asteroid belt and leave planetesimals with a significant total mass only within ~1 au. However, recent models of the formation of planetesimals by the streaming instability (Youdin and Goodman, 2005) showed that efficient planetesimal formation is possible only in regions with an enhanced dust/gas ratio due to radial pile-up of dust (Carrera et al., 2015; Yang et al., 2017;

Li and Youdin, 2021). Such a radial pile-up could have occurred near 1 AU due to the process of silicate sublimation and recondensation (Morbidelli et al., 2022), or due to the presence of a pressure bump (Ueda et al., 2021). This supports the idea of formation of terrestrial planets from a ring of planetesimals, without having to invoke a specific evolution of the giant planets as in previous works (Walsh et al., 2011; Izidoro et al., 2016; Clement et al., 2018; Lykawka and Ito, 2023).

Several papers have discussed the formation of terrestrial planets from a ring (Hansen, 2009; Nesvorný et al., 2021; Izidoro et al., 2022) but they started from a ring of planetesimals and of already-formed planetary embryos and neglected the gas disk phase and its effects on the migration of the embryos. Other works considered the growth of embryos form planetesimals (Morishima et al., 2010; Carter et al. 2015; Walsh & Levison 2019; Clement et al., 2020) but did not consider the case where planetesimals are confined in a ring. In our previous manuscript (Woo et al., 2023, hereafter Paper-I) we simulated the growth of planetary embryos during the gas-disk phase from a ring of 100 km planetesimals centred at 1 AU (with a Gaussian profile with σ=0.1 AU) and showed that migration and mutual scattering between embryos would cause them to spread from the original ring into a region of about 1 AU across. Such a radially spread distribution would unlikely lead to the formation of planets as massive as Earth and Venus even on a hundred-Myr timescale. Our results were similar to those of Deienno et al. (2019), although that work neglected the role of migration induced by the disk on the embryos.

Following Ogihara et al. (2018) and Brož et al. (2021), we also showed in Paper-I that a disk with a gas surface density peaked at 1 AU can prevent the radial spreading of planetary embryos by exerting on them convergent migration towards the density maximum. This leads to the formation of protoplanets with a few martian masses within the disk's lifetime in the Venus-Earth region (0.5 – 1.2 AU). Such a peaked density distribution can be the consequence of a viscosity enhancement due to magneto-rotational instability inward of 1 au (Flock et al., 2017), of enhanced disk winds (Suzuki et al., 2016; Ogihara et al., 2018) or even of the perturbations on the disk exerted by Jupiter (Izidoro, private comm.). However, the positive slope of the disk's density radial distribution within 1 AU should not be too steep, to avoid forming the full Earth within the disk lifetime, which would be inconsistent with the established geochronology (Kleine et al., 2009; Kleine and Nimmo, 2024).

The aim of this work is to extend the simulations of Paper-I and address the continuation of the assemblage of terrestrial planets in the post-gas phase. This requires extending the simulation to > 100 Myr. In turn, this means that we need to consider the contemporary evolution of the outer Solar system. The current giant planets' orbits and the architecture of the small body populations in the outer Solar system are evidence that the giant planets underwent a phase of instability, sometime after the dissipation of the gas disk (Tsiganis et al., 2005; Gomes et al., 2005; Morbidelli et al., 2007; Nesvorný and Morbidelli, 2012; Nesvorný, 2018; Clement et al., 2021a,b). The giant planet instability has been shown to be dynamically important in the inner Solar system, affecting the growth and orbital excitation of the terrestrial planets (Brasser et al., 2009, 2013; Roig et al., 2016; Kaib and Chambers, 2016; Clement et al., 2018, 2023; Nesvorný et al., 2021). Because the timing of the giant planet instability is not well known, but is bracketed between the dissipation time of the disk (Liu et al., 2022) and 100 My (Nesvorný et al., 2018; Mojzsis et al., 2019), in this study we enact it at different times: 5, 50 and 90 My after gas removal, and investigate how it affects the chronology of Earth formation, particularly the age distribution of the last giant impact that originated the Moon. We also consider the mass of the late veneer on Earth (Chou, 1978) and the final orbital excitation of the terrestrial planets. In the next section, we describe our simulation methods and set up. We present our results in Section 3. Section 4 discusses whether some key ingredient may still be missing in our simulations and Section 5 summarises our conclusions.

2. **Method**

We first perform simulations for 10 Myr with the GPU N-body code GENGA (Grimm and Stadel, 2014; Grimm et al., 2022), as in Paper-I. This GPU accelerated N-body integrator takes advantage of the large number of computing cores in GPU cards which can perform the same instructions on multiple threads in parallel. This speeds up both the mutual force calculation and the routines for handling close encounters, allowing the simulation of many interacting bodies.

Following Paper-I, we distribute the initial planetesimals within a narrow region according to a gaussian distribution function. The mean location of the ring $\mu$ is at 1 AU and the spread of the ring is $\sigma$ = 0.1 AU (i.e. ~68 % of the planetesimals are within $\mu \pm 1\sigma$ ). The total mass of the ring is 2.1 $M_{Earth}$, which is ~10 % more massive than the current system of terrestrial planets to account for the loss of mass through mutual scattering. The eccentricities and inclinations of the planetesimals are uniformly random in the range 0 < $e$ < 0.01 and 0° < $i$ < 0.5°. Their nodal angles and mean anomalies are uniformly random from 0° to 360°. Planetesimals are treated as super-particles with real initial sizes of ~100 km in diameter (Morbidelli et al., 2009; Shankman et al., 2013; Delbo et al., 2017, 2019), but with a mass ~1000 times their real mass (~2.6 x $10^{-4}$ $M_{Earth}$). In total we simulate about 8000 super-particles in a ring, all interacting with each other (gravitationally and collisionally). The details of how we treat gas drag, gravitational scattering as well as dynamical self-stirring of the super-particles are described in section 2.1 of Paper-I.

The planetesimals are placed in a disk of gas with an exponentially decaying surface density. The disk affects particles' evolution through gas drag (Adachi et al., 1976) and gravitational torques (Tanaka et al., 2002; Tanaka and Ward, 2004). As mentioned, the gas disk has a density peaked at ~ 1 AU to concentrate planetary embryos in the Earth-forming region via convergent migration (Ogihara et al., 2018; Brož et al., 2021). We name "convergent disk" a disk with such a structure. The surface density of the convergent disk at time $t$ is written as:

$$\sum_{gas}(r,t) = \sum_{gas,0} r^{-p} \exp(\frac{-t}{\tau_{decay}}),$$

(1)

where $r$ is the heliocentric distance from the Sun in AU, $\tau_{decay}$ = 1 Myr, $\Sigma_{gas,0}$ is assumed to be 1600 gcm$^{-2}$ and the exponent $p$ is negative for $r$<1 AU and positive for $r$>1 AU. The exact functional form of $p$ defines different convergent disks. We dub the disk with $p$= - 0.5 + 0.5{1 – tanh[(1 - $r$)/0.15]} the "shallow inner" disk, because the disk's surface density is less steep in region $r$ < 1 AU than the one proposed by Brož et al. (2021). This gas disk has been shown in the simulation of Paper-I to avoid forming the Earth too fast. For sake of comparison, Broz et al. did not consider a dissipating disk, but considered both an "early disk" with

$\Sigma_{gas,0}$=750g/cm³ and a "late disk" with $\Sigma_{gas,0}$=75g/cm³, while Ogihara et al. (2018) solved for the evolution of the gas surface density under the effect of viscosity and angular momentum removal in winds, resulting in an effective $\tau_{decay}$ of about 0.5 – 1 My and a surface density maximum moving over time.

We also test other radial profiles of the gas disk. First, we consider a case similar to the shallow inner disk, but with an even shallower inner slope, following equation (1) with $p$ = - 0.1 + 0.3{1 – tanh[(1 - r)/0.15]}. This is dubbed as the "shallower inner" disk. Then, to make the surface density peak at 1 AU flatter than in the shallow inner disk case, we test another radial distribution of the gas disk's surface density as:

$$\sum_{gas}(r,t) = \sum_{gas,0} \frac{5}{7r} atan((1.3r)^q) \exp\left(\frac{-t}{\tau_{decay}}\right),$$

(2)

where $q$ = 5 is used in the "flatpeak1" disk case, and $q$ = 3 in the "flatpeak2" case. The reason for assuming a convergent disk with a flatter peak is to verify whether it would form *two* massive planets in the Venus-Earth region (as opposed to one) more frequently than the "shallow inner" and "shallower inner" disk cases. Fig. 1 shows the initial surface density profiles of the different convergent disks tested in our study. Notice that the gas density in the flatpeak 1 & 2 cases is always below than that of the shallow inner and shallower inner disks. The difference can be as large as 30% at 1.5 au.

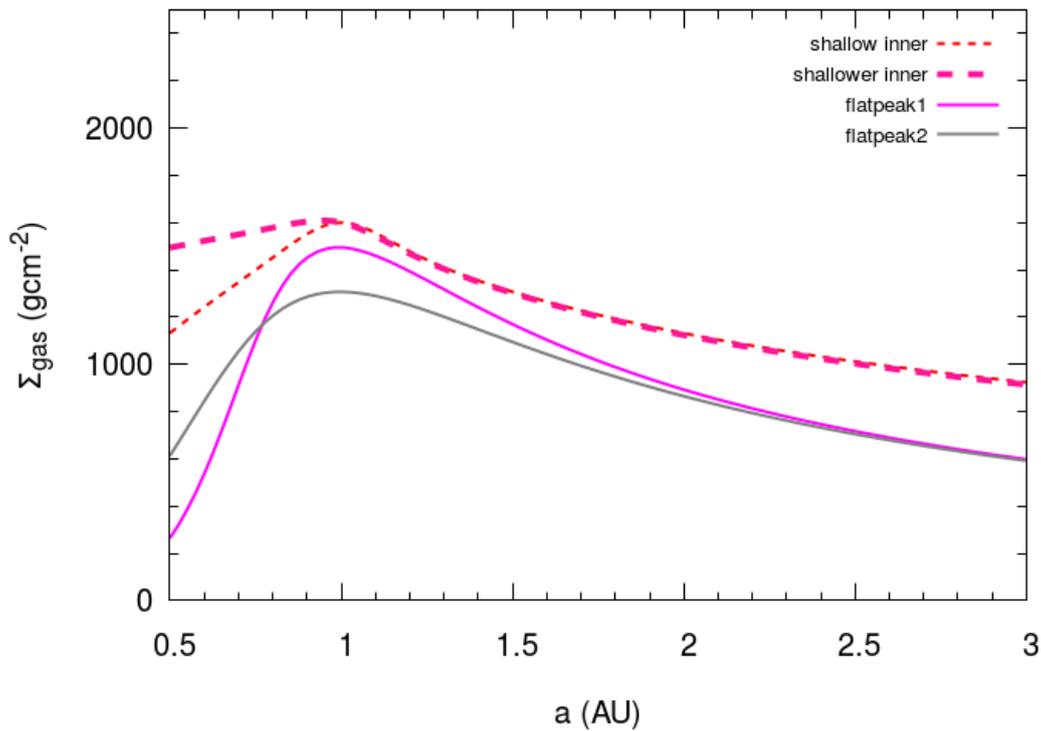

Figure 1 - The initial surface densities of the convergent gas disks tested in our study.

We perform at least 2 simulations for each gas disk for 10 Myr with GENGA. Some of the simulations for the "shallow inner" disk have been reported in Paper I, while the other three disk-cases are completely new. Jupiter and Saturn are placed on nearly circular orbits following O'Brien et al. (2006) (i.e. $a_j$=5.445 au, $e_j$=0.000916, $i_j$=0; $a_s$=8.178 au, $e_s$=0.000248 $i_s$= 0.0087 rad). Then we select the most successful simulation sets and continue them in the next stage. Admittedly, the choice of the most successful cases is rather subjective, given that the accretion process is not yet completed at 10 Myr: we look for simulations which form both an Earth not too fast (i.e. its most massive planet near 1 au is < ~0.5 $M_{Earth}$) and a Mars not too massive (< ~0.1 $M_{Earth}$).

We took our "best" end-states at 10 Myr and generated 10 sets of initial conditions from each of them, by slightly changing the velocity vectors of the planetesimals and embryos. We continued the simulations with the symplectic code Symba (Duncan et al., 1998), without any effects from the gas (the gas-disk is assumed be fully removed at 10My) and ignoring the self-interactions among planetesimals (i.e. among objects smaller than $3 \times 10^{-8}$ solar masses), given that the dynamics is now dominated by the presence of the planetary embryos. For the time-interval of +/- 5 My around the giant planet instability, we switched to the code iSyMBA (Roig et al., 2021), which is a modified version of SyMBA that interpolates from the 1-year resolution output of a previous simulation of the giant planet instability which resulted in final giant planet orbits very similar to the current ones. The advantage of the use of iSyMBA is that we don't waste CPU time on simulating instabilities that in most of the cases would lead to systems vastly different from our own, but simply re-inact a previously obtained successful simulation. (For more detail of iSyMBA, please refer to Roig et al., 2021). Because the path towards the current orbits during the instability is not unique, we considered two giant planet instability evolutions, dubbed *case1* and *case3* in Nesvorný et al. (2013) (see that paper for a detailed description of the giant plant dynamics in these two evolutions). The second provides less dynamical excitation in the inner Solar system than the former (Brasil et al., 2016). At 5 My after the instability, when the giant planets have reached final stable orbits similar to the present ones, we place the giant planets onto their exact current orbits and switch back to the use of the integrator Symba for the continuation of the simulations, till 210My. Table 1 summarizes the results of our simulations.

We perform simulations with three different instability timing: 15, 60, 100 Myr after the start of our GENGA simulations. Note that any chronological result in this paper is referred to $t$ = 0 of the GENGA simulation. The latter corresponds to the formation time of planetesimals in the ring, presumably within 1 My of the formation time of the calcium-aluminium inclusions, according to the chronology of formation of the parent bodies of iron meteorites (Kruijer et al., 2017). The total length of simulations is 210 Myr, in which 10 Myr is simulated with GENGA with a gas disk and 200 Myr is simulated with SyMBA and iSyMBA without a gas disk.

Our simulations do not include fragmentation during collisions at any stage. While during the gas-disk phase the relative velocities are low and collisions are merging events, after gas removal collisional velocities increase. After the giant planet instability, which results in a strong excitation of the terrestrial planet system, the assumption of perfectly merging collisions may become a crude approximation. Several works have explored the effects of generating debris during giant impacts (Kokubo and Genda, 2010; Chambers, 2013; Kobayashi et al., 2019; Crespi et al., 2021, 2024; Scora et al., 2020, 2022). However, the use of non-fully-accretional collisions is not yet a standard in terrestrial planet formation simulations, because there is no consensus on how fragments should be generated and modeled in a realistic way. The main aim of this work is to compare our results, stemming from the treatment of the initial growth of embryos form a self-interacting ring of planetesimals, with previous ones (Hansen, 2009; Broz et al., 2021; Nesvorny et al., 2021; Izidoro et al., 2022) which assumed an ad-hoc population of radially confined embryos. Because these previous works assumed all collisions to be mergers, we believe that it is better to adopt the same approximation for sake of comparisons.

Nevertheless, in Section 4.1 we will simulate the evolution of the systems that, on average, resulted in being the most excited during the giant planet instability with the code described in Scora et al. (2020), to estimate the total amount of debris mass and get an insight on their effectiveness in damping the dynamical excitation of the forming terrestrial planet system.

3. **Results**

   *3.1. Orbits and masses of the terrestrial planets*

We first present the final mass-distance distribution of the planets resulting from different gas disks and instability cases (Fig. 2). The overall distribution does not vary drastically with different gas disks or instability cases. Like previous ring models simulations (Morishima et al., 2008; Hansen, 2009; Walsh and Levison, 2016; Lykawka, 2020; Nesvorný et al., 2021; Izidoro et al., 2022), the distribution peaks at the Venus-Earth region (0.5 to 1.2 AU), with planets in Mercury (< 0.5 AU) and Mars region (1.4 < $a$ < 1.65 AU) having smaller masses on average. Notice that planets formed in the 1.2-1.4 AU region or beyond 1.65 AU are not considered successful reproductions of Earth or Mars, even if their mass is in the appropriate range.

Nevertheless, some differences in the results of simulations assuming different gas disks can be observed. Switching the disk from the "shallow inner" to the "shallower inner" clearly results in more massive planets forming in the Mercury region and less massive planets in the Venus region (0.5-0.9 AU). This is because more embryos can stay in the region < 0.5 AU if the disk's radial profile is shallower and, consequently, the outward migration is weaker. On the other hand, changing the disk from "shallow inner" to "flatpeak1" and "flatpeak2" results in more massive Mars. This could be due to a weaker convergent migration effect in the disk with a flatter peak at 1 AU but also to the overall reduced gas density mentioned before. There is no apparent reason for a systematic difference in Mars' mass between the "shallow inner" and "shallower inner" disk cases, so we believe that the apparent differences are just due to statistical sampling.

The statistical analysis of the simulations is presented in Table 1. In general, we formed 3 to 4 planets per simulation. In the following we evaluate the probabilities to form a Mars, a Mercury and a Venus-Earth pair independently.

The probability of forming a small planet in the Mars region is relatively low. Only 4 out of 12 series of simulations formed at least one Mars analogue with a correct mass of > 0.05 $M_{Earth}$ and < 0.2 $M_{Earth}$ (denoted *strict analogue* in Table 1). The most successful disk set-up is the "shallower inner" disk, with has about 50% chance of success of forming a Mars strict analogue with an average mass of 0.14 $M_{Earth}$, i.e. equivalent to the current mass of Mars. The other 3 disk cases have a success rate in forming Mars strict analogues of only 10 to 20 %, because the average mass of Mars generic analogues (i.e. in the 1.4-1.65 au interval irrespective of mass) is about 0.3 $M_{Earth}$. Part of the reason for such a low probability of forming a Mars analogue is that our definition of "analogue" requires that the planet resides in a narrow region (1.4 to 1.65 AU). We did this choice because planets formed inwards of 1.4 AU are typically more massive, given the overall mass-distance distribution of our resulting planets. This restricts also the upper limit to 1.65 au in order to place the real Mars at the center of the considered interval. As shown in Fig. 2 it is relatively easy to find low-mass planets farther than this upper limit, but this would artificially enhance the statistics of success.

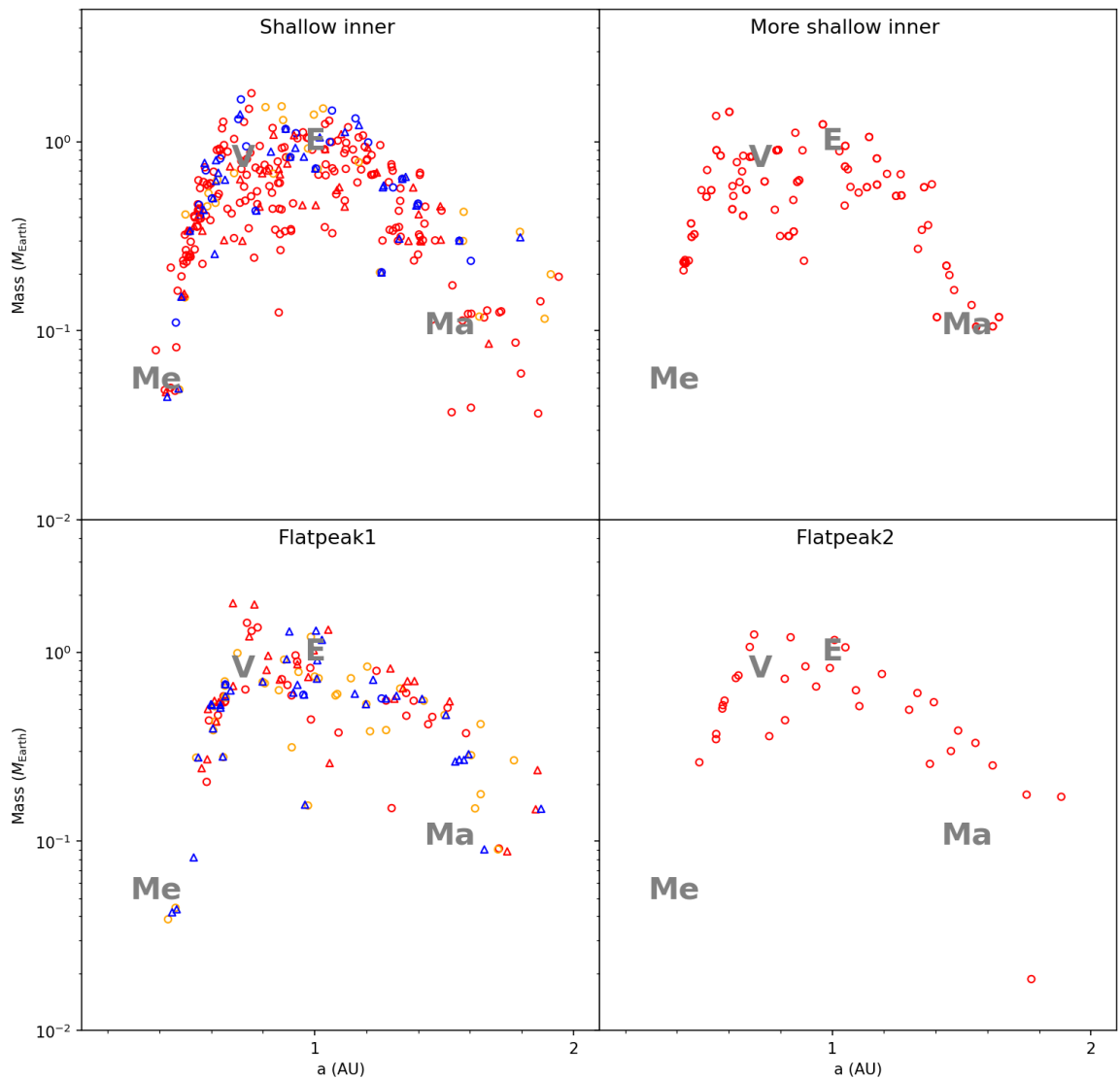

Figure 2 - The mass-distance distribution of the planets after 210 Myr of simulations in the four disk cases we considered. The circles refer to simulations performed enacting a giant planet instability of "case1", while triangles are for simulations performed with the "case3" instability. Red colour denotes data from simulations where the giant planet instability occurs at 15 Myr, yellow denote data from simulations with the instability at 60 Myr and blue denote data from simulations with the instability at 100 Myr. Only objects with mass > 0.01 $M_{Earth}$ are shown.

The largest number (16) of Mercury analogues with masses < 0.2 $M_{Earth}$ (denoted *strict analogue* in Table 1) are produced in the "shallow inner" disk simulations, partly because more simulations are performed for this gas disk profile. They are produced in 18% of the simulations, a success rate comparable to that of some sets of simulations of the "flatpeak1" series (see Table 1). Nevertheless, Mercury formation needs to be considered with caution, as fragmentation could play an important role in the inner Solar system due to high collisional velocities. It has been proposed that the high iron core-to-mantle mass ratio of Mercury could be a result of giant impact(s) (Benz et al., 1988; Asphaug et al., 2006; Svetsov, 2011; Chau et al., 2018; Clement et al., 2023), or collisional erosion of its precursor planetesimals (Hyodo et al., 2021). Our simulations do not include fragmentation, and hence may not be able to make a reliable conclusion on Mercury's formation.

The simulations form Venus-Earth pairs relatively well. Parameter $f_g$ in Table 1 denotes the fraction of simulations that form exactly two planets with mass > 0.3 $M_{Earth}$ in the region 0.5 to 1.2 AU. Notice that $f_g$ is always larger than 0.5 and is marginally higher in the flatpeak1 series. Increasing the mass limit to 0.5 $M_{Earth}$ would decrease $f_g$ only marginally (about 0.5 for the shallow disk case). Table 1 also shows the mean separation of Venus and Earth ($\Delta a_{Venus-Earth}$) and the mean orbital eccentricity/inclination of Venus-Earth ($<e,i>_{Venus-Earth} = (e_{Venus} + i_{Venus} + e_{Earth} + i_{Earth})/4$, where $e$ and $i$ are eccentricity and inclination respectively), as defined in Nesvorny et al. (2021). In about half of the simulations, the 1-standard deviation of the distributions of $\Delta a_{Venus-Earth}$ and $<e,i>_{Venus-Earth}$ overlap with the values of the current real planets, showing that the ring model is successful in forming Venus-Earth analogues with good dynamical properties.

Concerning the dynamical properties of the terrestrial planets, Table 1 also reports the median, 25th and 75th percentile of the radial mass concentration parameter:

$$S_c = \max\left(\frac{\sum_k m_k}{\sum_k m_k (\log(a/a_k))^2}\right),$$
(3)

where $m_k$ and $a_k$ are the mass and semi-major axis of each planet produced in the simulation, regardless of whether it "reproduces" a real planet or not. The median, 25th and 75th percentile of the normalised angular momentum deficit:

$$\text{AMD} = \frac{\sum_k m_k \sqrt{a_k}[1 - \sqrt{(1-e_k^2)}\cos I_k]}{\sum_k m_k \sqrt{a_k}} / \text{AMD}_{current},$$
(4)

which measures the eccentricity and inclination excitation of a system, are also shown in Table 1. In formula (4) $e_k$ and $I_k$ are the eccentricity and inclination of each planet and $\text{AMD}_{current} = 0.0018$ is the current AMD of the terrestrial planets. Our simulations in general, fail to reproduce the high $S_c$ of the current Solar system, because many simulations form too massive Mars or Mercury, even if the overall mass distribution has the correct pattern. In some cases, for example in the suite of simulations with a "flatpeak1" disk and an instability of type "case 1" at 15 My, we obtain a high median $S_c$, but this is because we often form less than 3 planets (in 3 out of 10 runs). Indeed, $S_c$ can be very high (> 100) if there are only two planets, both located in the Venus-Earth region, or a single massive Earth at ~1 AU and a tiny Mars at ~1.5 AU, as in some of our flatpeak1 simulations.

Table 1 - Statistics of each simulation set. The parameters from left to right are (1) the number of simulations performed to 210 Myr ($N_{sim}$); (2) the average number of planets with mass > 0.01 $M_{Earth}$ in each simulation ($N_{avg}$); (3) the average total mass of the planets in the terrestrial planet region ($a < 2$ AU, $a$ is semi-major axis) (4) **mean** number of Mars analogues ($1.4 < a < 1.65$ au) per simulation and of strict Mars analogues with mass > 0.05 $M_{Earth}$ and < 0.2 $M_{Earth}$ per simulation ($N_{Mars}/N_{Mars\_strict}$); (5) the averaged mass of Mars ($M_{Mars}$); (6) **mean** number of Mercury analogues ($a < 0.5$ AU) per simulation and of strict Mercury analogues with mass < 0.2 $M_{Earth}$ per simulation ($N_{Mercury}/N_{Mercury\_strict}$); (7) the averaged mass of Mercury ($M_{Mercury}$); (8) fraction of simulations which produce exactly two planets with mass > 0.3 $M_{Earth}$ in the Venus-Earth region ($0.5 < a < 1.2$ AU) ($f_g$); (9) the average separation of Venus and Earth in simulations producing exactly two planets in the Venus-Earth region ($\Delta a_{Venus-Earth}$); (10) the mean orbital eccentricity/inclination of Venus-Earth in simulations with exactly two planets in the Venus-Earth region ($<e,i>_{Venus-Earth}$); (11) the median, 25th and 75th percentile of the radial concentration parameter ($S_c$, see equation (3)); (12) the median, 25th and 75th percentile of the normalized angular momentum deficit of the system (AMD; see equation (4)). The data for the current Solar system are reported in the first row, for comparison. Unless specified, the uncertainty of the data reported in the table denotes one standard deviation from their mean.

| Gas disk | Instability cases | Instability time | $N_{sim}$ | $N_{avg}$ | $Mass_{avg}$ | $N_{Mars}/N_{Mars\_strict}$ | $M_{Mars}$ | $N_{Mercury}/N_{Mercury\_strict}$ | $M_{Mercury}$ | $f_g$ | $\Delta a_{Venus-Earth}$ | $<e,i>_{Venus-Earth}$ | $S_c$ | AMD |
|---|---|---|---|---|---|---|---|---|---|---|---|---|---|---|
| Current Solar system | | | … | 4 | 1.98 | 1/1 | 0.11 | 1/1 | 0.06 | … | 0.28 | 0.027 | 89.9 | 1 |
| Shallow inner | Case1 | 15 Myr | 50 | 3.5 | 1.94 ± 0.09 | 0.32/0.1 | 0.29 ± 0.20 | 0.22/0.14 | 0.15 ± 0.09 | 0.54 | 0.36 ± 0.10 | 0.053 ± 0.058 | $57.6^{60.9}_{51.0}$ | $1.05^{4.73}_{0.53}$ |
| | | 60 Myr | 10 | 2.9 | 1.88 ± 0.19 | 0.3/0.1 | 0.28 ± 0.15 | 0.2/0.2 | 0.10 ± 0.07 | 0.6 | 0.38 ± 0.08 | 0.094 ± 0.090 | $59.2^{64.7}_{53.5}$ | $5.37^{19.2}_{1.64}$ |
| | | 100 Myr | 10 | 2.6 | 1.92 ± 0.20 | 0.2/0 | 0.27 ± 0.05 | 0.1/0.1 | 0.11 | 0.5 | 0.41 ± 0.10 | 0.086 ± 0.052 | $55.8^{65.6}_{53.2}$ | $4.95^{8.06}_{2.54}$ |
| | Case3 | 15 Myr | 10 | 3.9 | 1.98 ± 0.04 | 0.5/0 | 0.35 ± 0.07 | 0.3/0.3 | 0.12 ± 0.06 | 0.5 | 0.28 ± 0.09 | 0.039 ± 0.019 | $57.9^{59.9}_{54.7}$ | $0.67^{1.83}_{0.56}$ |
| | | 100 Myr | 10 | 3.2 | 1.99 ± 0.06 | 0.1/0 | 0.30 | 0.3/0.3 | 0.082 +/- 0.060 | 0.9 | 0.36 ± 0.12 | 0.085 ± 0.065 | $56.4^{63}_{50.1}$ | $2.90^{8.78}_{0.85}$ |
| Shallower inner | Case1 | 15 Myr | 20 | 3.8 | 2.00 ± 0.03 | 0.65/0.55 | 0.14 ± 0.04 | 0.65/0 | 0.28 ± 0.10 | 0.60 | 0.37 ± 0.11 | 0.070 ± 0.030 | $44.8^{45.4}_{40.4}$ | $1.92^{4.03}_{1.12}$ |
| Flatpeak1 | Case1 | 15 Myr | 10 | 3.0 | 1.85 ± 0.13 | 0.4/0 | 0.44 ± 0.06 | 0/0 | NIL | 0.60 | 0.28 ± 0.04 | 0.058 ± 0.034 | $61.7^{75.8}_{57.7}$ | $2.86^{4.69}_{1.30}$ |
| | | 60 Myr | 10 | 3.8 | 1.94 ± 0.06 | 0.6/0.2 | 0.34 ± 0.16 | 0.2/0.2 | 0.042 ± 0.004 | 1 | 0.34 ± 0.07 | 0.063 ± 0.039 | $59.3^{67.0}_{51.3}$ | $1.10^{5.89}_{0.78}$ |
| | | 100 Myr | 10 | 3.2 | 1.92 ± 0.09 | 0/0 | NIL | 0/0 | NIL | 0.7 | 0.33 ± 0.07 | 0.060 ± 0.020 | $62.5^{69.6}_{55.7}$ | $2.17^{3.57}_{1.42}$ |
| | Case3 | 15 Myr | 10 | 2.7 | 1.91 ± 0.06 | 0.1/0 | 0.55 | 0/0 | NIL | 0.5 | 0.34 ± 0.08 | 0.051 ± 0.031 | $57.2^{64.1}_{55.5}$ | $2.55^{5.40}_{0.82}$ |
| | | 100 Myr | 10 | 3.7 | 1.95 ± 0.07 | 0.6/0 | 0.35 +/- 0.13 | 0.2/0.2 | 0.043 +/- 0.001 | 0.8 | 0.35 ± 0.05 | 0.043 ± 0.023 | $60.4^{67.0}_{55.5}$ | $1.06^{2.53}_{0.59}$ |
| Flatpeak2 | Case1 | 15 Myr | 10 | 3.3 | 1.90 ± 0.08 | 0.4/0 | 0.32 ± 0.06 | 0.1/0 | 0.26 | 0.50 | 0.40 +/- 0.07 | 0.050 ± 0.030 | $51.3^{55.4}_{49.4}$ | $2.43^{4.47}_{1.52}$ |

The normalised AMD is in general better reproduced than $S_c$. 4 out of 12 sets have a median AMD less than 1.5 $AMD_{current}$ and half of the simulation sets have the 25th percentile AMD less than 1 $AMD_{current}$, which indicates that it is not statistically unlikely to form a terrestrial planet system with a low enough AMD, even accounting for the giant planet instability. However, we observe a trend where the AMD of a system increases as the timing of giant planet instability is delayed, especially in the shallow inner disk case, where we performed more simulations. Fig. 3 compares the AMD evolutions in two (typical) simulations, where the giant planet instability occurred at 15 Myr, and 100 Myr respectively. We observe that the giant planet instability excites the AMD of the system of embryos in both cases. If this happens late, the AMD excitation is stronger because the embryos are more massive, and they self-excite more strongly during the mutual close encounters phase that follows the instability. Moreover, fewer planetesimals remain in the simulation, capable of damping the terrestrial planets system by dynamical friction (Wetherill and Stewart, 1989).

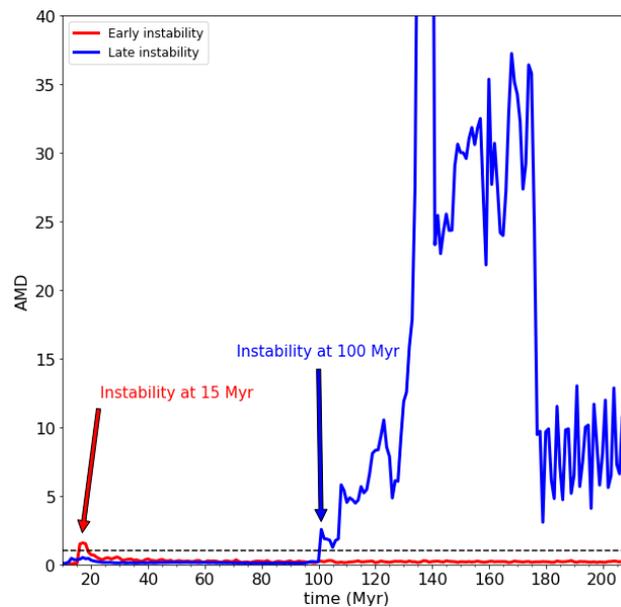

Figure 3 - AMD evolution for a simulation enacting the giant planet instability at 15 Myr (early instability, red) and one enacting the same instability at 100 My (late instability, blue) simulation. Both simulations start from the same initial conditions, adopt the the "shallow inner" disk profile and the instability evolution of type "case 1".

### 3.2. *The last giant impact and the late veneer mass on Earth*

In this section, we compare the growth of the Earth-mass planets (M>0.5 $M_{Earth}$) and the timing of their last giant impact observed in our simulations with chronological constraints discussed in the literature. For sake of statistics, we don't distinguish between the innermost (Venus) or outermost (Earth) of the two major planets produced in the simulations, considering that what happened on Earth could have happened on Venus and vice-versa.

Fig. 4 shows the growth history of all the Earth-sized planets obtained in our simulations. We find that Earth-sized planets tend to complete their formation earlier when the instability occurs earlier. Most of the Earth-sized planets reached 80% of their final mass within 50 Myr in the simulations where the giant planet instability was enacted at 15 Myr (red lines). This seems incompatible with the Hf-W chronology, which bounds how fast Earth growth could have been, as illustrated by the pink region in Fig. 4. These are the "forbidden region" of Earth's growth (Rudge et al., 2010), assuming that ~40% of the impact's core equilibrates with the whole mantle of the target in each collision. The pink region shrinks/expands assuming less/more equilibration, but 40% is the equilibration fraction that best reconciles the Hf-W and Pb-Pb chronometers (see Rudge et al., 2010). Most of the Earth size planets have a growth curve crossing the pink area in the early instability case, indicating that these planets should have in the end an overabundance of radiogenic $^{182}$W in their mantle with respect to the real bulk silicate Earth (BSE). In contrast, many simulations enacting a late giant planet instability do not cross the forbidden region; these cases appear more compatible with the Hf-W chronology. A detailed model of the evolution of $\varepsilon^{182}$W during the growth of Earth in these simulations will be developed in a forthcoming work, following Rubie et al. (2024).

Nearly all Earth-sized planets undergo at least one giant impact, which is defined as a collision between two objects both having at least a lunar mass (~0.01 $M_{Earth,}$). These giant impacts are indicated by big jumps in mass in the growth histories of Earth-sized planets in Fig. 4. These jumps occur mostly in the first 50 Myr for the early instability case (red lines in Fig. 4), while many more occur around or beyond 100 Myr in the later instability cases (orange and blue lines in Fig. 4). The last giant impact being usually thought to be related to the formation of the Moon, its timing can be related to the Moon chronology. The left panel of Fig. 5 shows the cumulative distribution function (CDF) of the timing of the last giant impact on the Earth-sized planets. It is obvious that the last giant impact tends to occur earlier when the giant planet instability happens earlier. 50 % of the last giant impacts occur not later than 25 Myr after the giant planet instability. In the cases where the giant planet instability is enacted at 60 or 100 Myr, 35% of the Earth or Venus analogs suffer a last giant impact before the instability; of the remaining 65% of Earth or Venus analogs, half suffer a last giant impact in the 50 Myr following the instability. This is because the giant planet instability excites the eccentricities of the orbits of the embryos in the terrestrial region (see Fig. 3), causing them to cross each other and leading to mutual impacts. Thus, the giant planet instability controls the chronology of the giant impacts, including the timing of the last giant impact, i.e. of the Moon-forming event.

The right panel of Fig. 5 shows the CDF of the impactor to target mass ratio of the last giant impact on the Earth-sized planets. 90% of the giant impacts have $\gamma > 0.1$ and more than 50% of them have $\gamma > 0.3$. This is due to several similar size embryos with ~0.3 - 0.5 $M_{Earth}$ forming in the Venus-Earth region in the first 10 Myr. Earth's formation results from collisions of these similar size embryos.

The high frequency of giant impacts with $\gamma > 0.3$ is a characteristic feature of our model. It has been shown in previous studies that large scale impacts are rare in the Classical model (Kaib and Cowan, 2015; Woo et al., 2022) and the Grand Tack model (Jacobson and Morbidelli, 2014). These studies assume that hundreds of small embryos exist in the disk initially. Hence there are always small embryos of lunar to martian mass that remain in the system until the late stage of a simulation and thus available for a last collision with the proto-Earth. Instead, in our ring model simulation, embryos grow self-consistently from the ring of planetesimals and lunar to martian size embryos are rare in the Earth-region after 10 Myr.

The canonical model of Moon formation assumes a collision between a Mars-sized impactor and the proto-Earth. The impactor to target mass ratio, $\gamma$ is thus ~0.1. Our results suggest that Theia was most likely larger than 2 $M_{Mars}$ with $\gamma > 0.3$. A larger Theia colliding with Earth should be more favourable in producing a similar isotopic composition in Earth and the Moon, due to either the emplacement of a higher portion of Earth's mantle ending in the proto-lunar disk (Canup, 2012) or the evaporation of part of the Earth and the production of a synesthia (Lock et al., 2018).

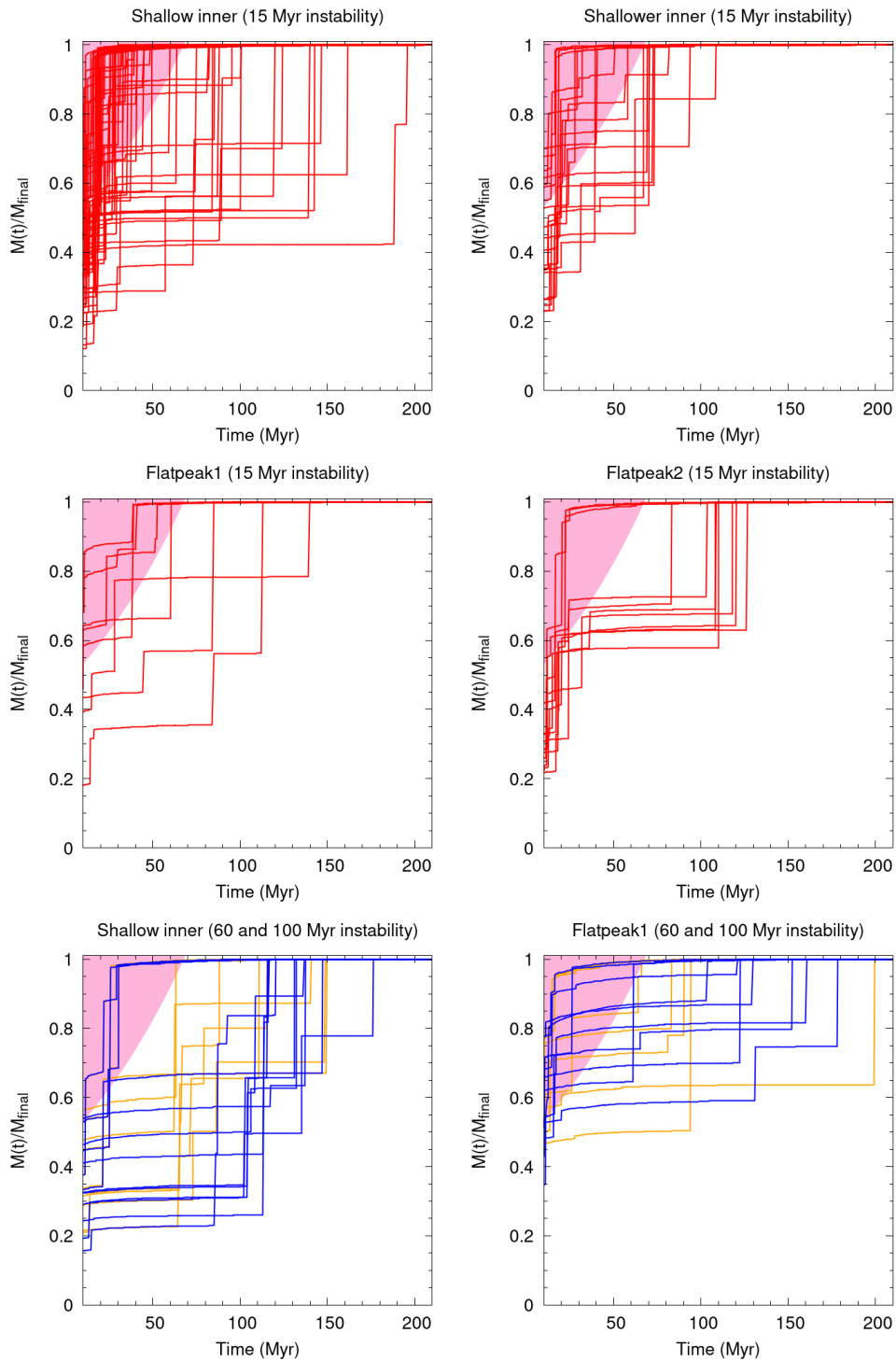

Figure 4 - Growth history of the Earth-size planets in our simulations starting with four different gas disks profiles. The red, orange, blue lines show results for the simulations enacting the instability at 15 Myr, 60 Myr and 100 Myr, respectively The pink region denotes the forbidden region according to Hf-W chronology (Rudge et al., 2010). This means that any planet whose growth history crosses this region would have in the end a too large value of $\varepsilon^{182}W$ assuming the equilibration factor $k$ = 0.4 (i.e. ~40% of the impactor's core equilibrates with the whole target's mantle in each collision). Note that the abscissa starts at 10 Myr.

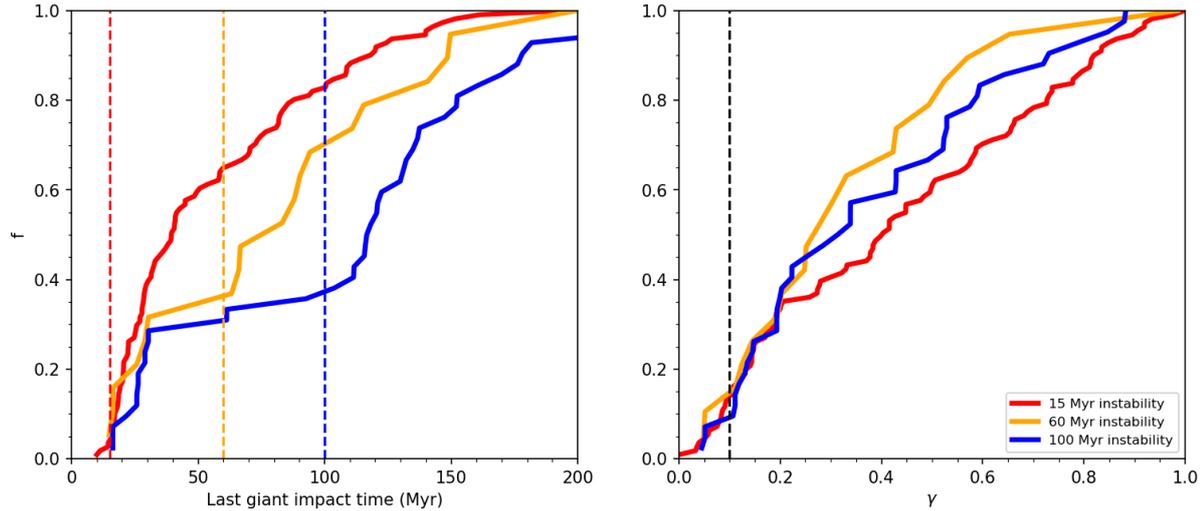

Figure 5 - The cumulative distribution function (CDF) of the timing of the last giant impact (left) and of the impactor-to-target mass ratio ($\gamma$) in the last giant impact (right) for Earth-sized planets. The red, orange and blue colours correspond to simulations enacting the giant planet instability at 15 Myr, 60 Myr and 100 Myr, respectively. These times are indicated by the vertical dashed lines of the same colours. The black vertical dashed line in the right panel denotes $\gamma = 0.1$, which is the impactor-to-target mass ratio in the canonical Moon-forming giant impact (Canup and Asphaug, 2001).

The exact timing of the Moon-forming giant impacts remains elusive. In general, there are three ways to date the formation of the Moon: dating the crystallisation of the lunar magma ocean (Gaffney and Borg, 2014; Borg et al., 2011, 2019; Barboni et al., 2017), the formation of the lunar core (Thiemens et al., 2019; Kruijer et al., 2021) or the volatile loss from the Moon (Halliday, 2008; Mezger et al., 2021; Borg et al., 2022; Connelly et al., 2022). Each event is dated by different radioactive chronometers, often providing conflicting ages. Combining the model ages of all these three events, Kleine and Nimmo (2024) concluded that most likely the formation of the Moon occurred at a time comprised between 75 to 150 Myr after the condensation of the first solids of the Solar system, that we use as a reference hereafter.

In our simulations, we observe last giant impacts occurring between ~20 and 200 My, so embracing the time-range discussed above. However, we can use correlations between the last giant impact time and the amount of late veneer on Earth or the final AMD of the terrestrial planets to infer which last giant impact time is more likely.

The late veneer traces the amount of material accreted by Earth after the end of its core formation, which most likely coincides with the Moon-forming event (Morbidelli and Wood, 2015). The late veneer mass is constrained by the abundance of highly siderophile elements (HSEs) in the bulk silicate Earth and is $(4.8 \pm 1.6) \times 10^{-3}$ $M_{Earth}$ (Walker, 2009). Fig. 6 shows the amount of late veneer, relative to the planet's final mass, versus the timing of the last giant impact for each Earth-sized planet in our simulations. There is a clear anti-correlation between the late veneer mass and the timing of the last giant impact. This is expected as there are fewer planetesimals available for the late veneer in the later stage of the simulation (Jacobson et al., 2014).

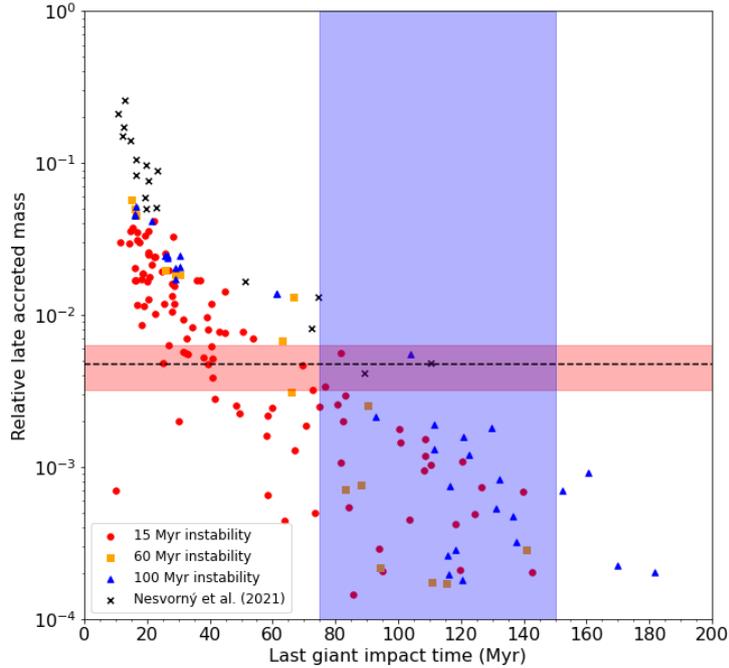

Figure 6 - The mass of the late veneer, measured relatively to planet's final mass, versus the time of the last giant impact on the Earth-sized planets of our simulations. We also plot the same quantities for the best 10 simulations of Nesvorny et al. (2021) as black crosses. The blue shaded region indicates the range of possible times of the Moon-forming event inferred from combining several lunar chronometers (Kleine and Nimmo, 2024). The red shaded region indicates the relative amount of late veneer accreted by Earth as deduced from the abundance of highly siderophile elements in the bulk silicate Earth (Walker, 2009).

Our results are nevertheless quantitatively different from those of Jacobson et al. (2014), who considered Grand Tack and Classical simulations. With a similar last giant impact time, the late veneer mass in Jacobson et al. (2014) is always nearly an order of magnitude higher than the late veneer mass obtained in our simulations. They conclude that the Moon-forming giant impact should have occurred at ~100 Myr after the birth of the Solar system in order to match the late veneer mass recorded in the BSE, whereas in our study a giant impact at 100 Myr most likely results in late veneer mass of only a few $10^{-4}$ $M_{Earth}$. We believe that the main reason for such a difference is not due to the confinement of the initial conditions in a ring or in an extended disk but is due to the different setup of the masses of the initial objects. The simulations considered in Jacobson et al. 2014 started from an assumed population of already-grown embryos embedded in a planetesimal disk, which were expected to exist at the end of the gas-disk phase. Instead, our simulations grow the embryos self-consistently from a ring of planetesimals during the gas-disk phase. Therefore, planetesimals are consumed faster in our simulations than in the simulations considered by Jacobson et al. (2014). Thus, at the end of the gas-disk phase the ratio between the total mass in embryos and in planetesimals is significantly larger in our simulations than assumed as initial conditions by Jacobson et al.

To confirm this interpretation, we also plotted in Fig. 6 the late veneer masses of the Earth-sized planets from ring model simulations of Nesvorný et al. (2021). The values are again a few times to an order of magnitude higher than those of our simulations, for the same giant impact time. Indeed, Nesvorný et al. (2021) also did not grow the embryos from planetesimals. but assumed the existence of embryos together with a disk of planetesimals comprising a total of 2 $M_{Earth}$ masses. For comparison, only < 0.2 $M_{Earth}$ remain in the planetesimals after the gas disk dissipation in our simulations. We stress that this is not due to an insufficient mass in our initial planetesimal rings. In 10 out of the 12 series of simulations that we performed (Table 1), the final mass comprised in the terrestrial planets exceeds 1.9 Earth masses. We may increase the initial ring mass by 10% but no more. This would probably increase the late veneer mass by the same relative amount, which would not change Fig. 6 significantly. The small amount of total mass in planetesimals at the end of the gas-disk phase is due to the rapid incorporation of the planetesimals in the planetary embryos during the gas-disk phase, a process never accounted for before. Thus, we expect that the late veneer masses recorded in our simulations are more realistic.

The correlation between the late veneer mass and the last giant impact time shown in Fig. 6 by our simulations suggests that reproducing statistically the real late veneer mass requires that the last giant impact occurred between 40 and 60 Myr. This is apparently excluded by the analysis of radioactive chronometers of Moon formation. Of the time interval suggested by Moon-formation chronology, only at the lower end (~75-85 Myr) we observe some simulations (about ~1/4 of the total) resulting in the correct late veneer mass. From the left panel of Fig 5, a last giant impact has a ~8% probability to occur in the 75-85 My range if the giant planet instability was early, ~4% if it was at 60 Myr and ~1% if it was at 100 Myr. However, given the uptick of the cumulative distribution curves just after the instability, the probability could have been significantly larger (possibly up to ~15%) if the instability had occurred around 70-75 Myr.

We also find a negative correlation between AMD and late veneer mass (Fig. 7, left panel). It is understandable because a small late veneer mass implies an anaemic population of left-over planetesimals and therefore a weak dynamical friction to damp the excitation of the final terrestrial planet system. Because we have three constraints (time of the last giant impact, AMD and late veneer mass), to verify whether there are planets that can approximately satisfy all of them and see in which scenario they are produced, we plot on the right panel of Fig. 7 the time of the last giant impact as a function of the normalized AMD of the final terrestrial planet system *selectively* for those planets with $2\times10^{-3}$ < LVM < $2\times10^{-2}$ Earth masses. This range is less restrictive than the one depicted by the red-coloured band in the left panel but allows to increment statistics. If we restrict ourselves to the Moon formation time interval 75-150 My, we are left with only 4 planets with the correct normalized AMD ~1: three from early instability simulations and one from a late instability simulation. If we decrease the lower limit of Moon formation time to 50 My, we get 5 additional good planets, 3 from the 60 My instability, 1 from the early instability and 1 from the late instability runs. There would be many more successful cases if the Moon could have formed before 40 My, but this would get in conflict with the two-stage Hf-W model of core formation of Earth (Kleine and Walker, 2017). Notice that there are no planets with last giant impact after ~100My in the panel. This is because all planets suffering a later Moon-forming impact receive a LVM <$2\times10^{-3}$ Earth masses, as also visible in Fig. 6.

It is undoubtedly difficult to match all the constraints set by geochemistry, chronology and dynamical properties. Nevertheless, the fact that the various trends that we have discussed have a non-null intersection gives hope that the real terrestrial planet system could be reproduced if the number of simulations is large enough. This work is devoted to a qualitative exploration of the scenario of terrestrial planet formation from a ring and, consequently, has limited statistical power. We will improve the statistics in a future work once it will become clear to us which parameters are most favourable.

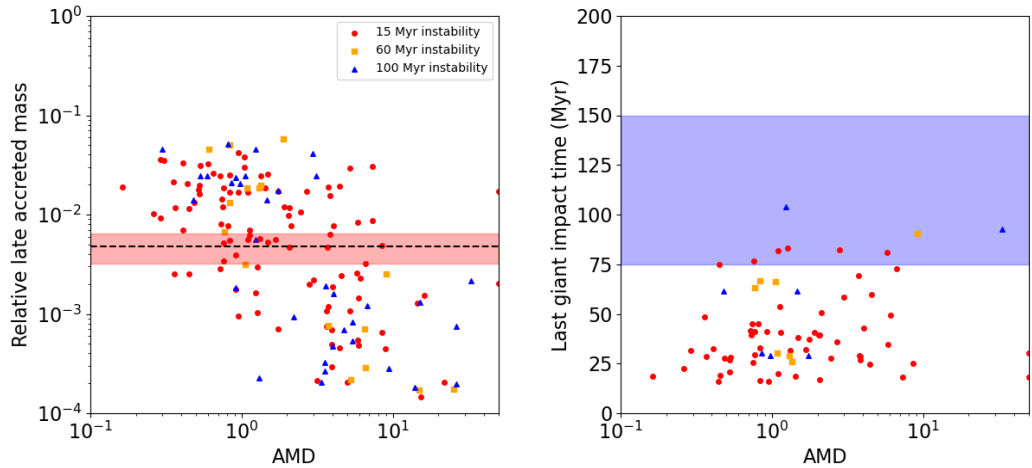

Figure 7 – Left: the late veneer mass of each Earth-sized planet versus the planet systems' final AMD. Right: the time of last giant impact vs. the final AMD for the planets with Late Veneer mass between $2 \times 10^{-3}$ and $2 \times 10^{-2}$ Earth masses. This is a bit broader than the red shaded band in the left panel, to improve statistics. The shaded regions are the same as those in Fig 6.

### 4. Discussion: is something still missing?

The results presented in the previous section show that there is a tension between obtaining in simulations a late enough last giant impact on Earth, a large enough late veneer mass and a small enough AMD. There is an intersection of these three constraints, but it is small, mostly limited to a minority of simulations featuring the last giant impacts in the 75-85 Myr interval. The situation would improve if the Moon age could be revised to an earlier time, 50 or 60 My after the beginning of the Solar system. It would also improve if we could find an additional source increasing the amount of late veneer and/or decreasing the final AMD. We discuss below some ideas, but which unfortunately do not appear very promising.

One idea to increase the late veneer mass on Earth is to have more planetesimals accreted during the final stage of planet formation. Our setup of the ring model has only accounted for the first generation planetesimals that formed in the ring. These planetesimals are probably the parent bodies of the achondrite meteorites of non-carbonaceous isotopic composition, formed in the first few ~0.1 Myr after the first solids of the Solar system. They are very efficiently consumed in the process of embryo's growth in the first few My, when gas drag is still strong. But we know that at 2-4 My chondritic planetesimals formed, possibly in the ring, possibly in the asteroid belt, or in both. At that time embryos were already formed, and gas drag was weaker. Thus, these later planetesimals would have behaved more similarly to those studied in simulations that start with already formed embryos at the end of the gas-disk phase. So, they could have potentially delivered more mass later, as in Jacobson et al., 2014. However, the simulations of Nesvorny et al. (2021) showed that only a small fraction of the planetesimals indigenous of the asteroid belt collide with the terrestrial planets, so it is unclear whether they can increase the late veneer mass by a significant factor.

Carter and Stewart (2022) proposed that the late veneer was carried by leftover differentiated planetesimals which had lost part of their silicates in mantle-stripping collisions. These objects would therefore deliver more HSEs per unit mass, thus lowering the total late veneer mass with respect to that classically deduced from the HSE abundance in the BSE assuming a chondritic composition of the projectiles. Although this idea is valid, in the framework of the simulations presented here it would not change our results. This is because our simulations, which don't feature mantle stripping collisions from planetesimals, preserve their condritic concentration of HSEs and therefore the mass delivered in the simulations after the last giant impact can be directly compared with the classic late veneer mass.

Another potential possibility to increase the late veneer mass is to include the ejection of core debris during giant impacts. Genda et al. (2017) showed that, under some impact conditions, the core of a lunar mass projectile can be fragmented and ejected into heliocentric orbits. These iron-rich fragments can then be re-accreted by the Earth and remain in its mantle. Even without invoking the ejection of core debris Korenaga and Marchi (2023) also showed that a lunar-mass embryo could deliver a significant fraction of its metal to the base of Earth's mantle. Thus, a single lunar mass embryo could account for all late veneer on Earth given that the metal is rich in HSEs. Unfortunately, we don't find a significant number of lunar mass embryos impacting the terrestrial planets in our simulations. But if a similar fate could happen to a smaller fraction of the core of Theia, this could be a sufficient source of late veneer.

Unlike the fragments originated from the impactor's core the fragments ejected from the mantle of the target or the projectile would not contribute to the late veneer because their HSEs concentrations are too low. These mantle-originated fragments, however, could help damping the dynamical excitation of the terrestrial planet orbits through dynamical friction, reducing the final AMD of the system (Kobayashi et al., 2019). To test this idea, we performed some simulations accounting for debris ejection during giant impacts, reported below.

### 4.1 Debris-generating collisions

We repeated the simulations of the 20 systems affected by a late giant planet instability of type "case 1", obtained from "flatpeak1" and "inner shallow" disk profiles (see Table 1), using a version of the SyMBA N-body code that accounts for non-accretional impacts (Duncan et al. 1998; Scora et al 2020). The code uses the analytic collision prescriptions of Stewart & Leinhardt (2012), Leinhardt & Stewart (2012) and Genda et al. (2012) to determine collisional outcomes based on the impact parameters of giant collisions within the simulations. These collisional outcomes include the creation of debris particles, which are tracked throughout the simulation as interacting with other embryos, but not with other debris particles. To keep the code from slowing down too much, the debris particles have a minimum mass and maximum number of particles, so they act as tracers for what would be many smaller debris particles. For further details on how the code handles collisions, see Scora et al. (2020).

The new simulations start at the end of the iSymba runs implementing the giant planet instability and cover an additional timespan of 100 to 300 Myr. Figure 8 shows the initial and final AMD of each system, the duration of each simulation and the total mass of generated debris.

The systems have a post giant-planet instability AMD ranging from 1.5x to 7x the current AMD of the terrestrial planets. For most of them the final AMD is smaller than the initial one and for about a third of them the final AMD is less than 1.5x the current AMD. However, this evolution is not due predominantly to debris generation, but rather to the effect of merging events and the interaction with the original planetesimals. In fact, the total mass in generated debris is always small, never exceeding ½ of an Earth mass and in most cases smaller than 0.1 Earth masses. There is no apparent correlation between AMD decrease and generated debris mass. The statistics on the distribution of the final AMDs is very similar to that of the original simulations treating all collisions as perfect mergers. The median final AMD was 3.55x the current one and is 2.5x now. The AMD was larger than 10x the current one in 17% of the original runs, whereas the same is true in 10% of the new runs. Thus, there is an effect due to debris generation, but it is small. Debris generation is unlikely to change the statistical correlation depicted in Fig. 7 showing an excessive AMD for cases with late instabilities and late Moon-forming events.

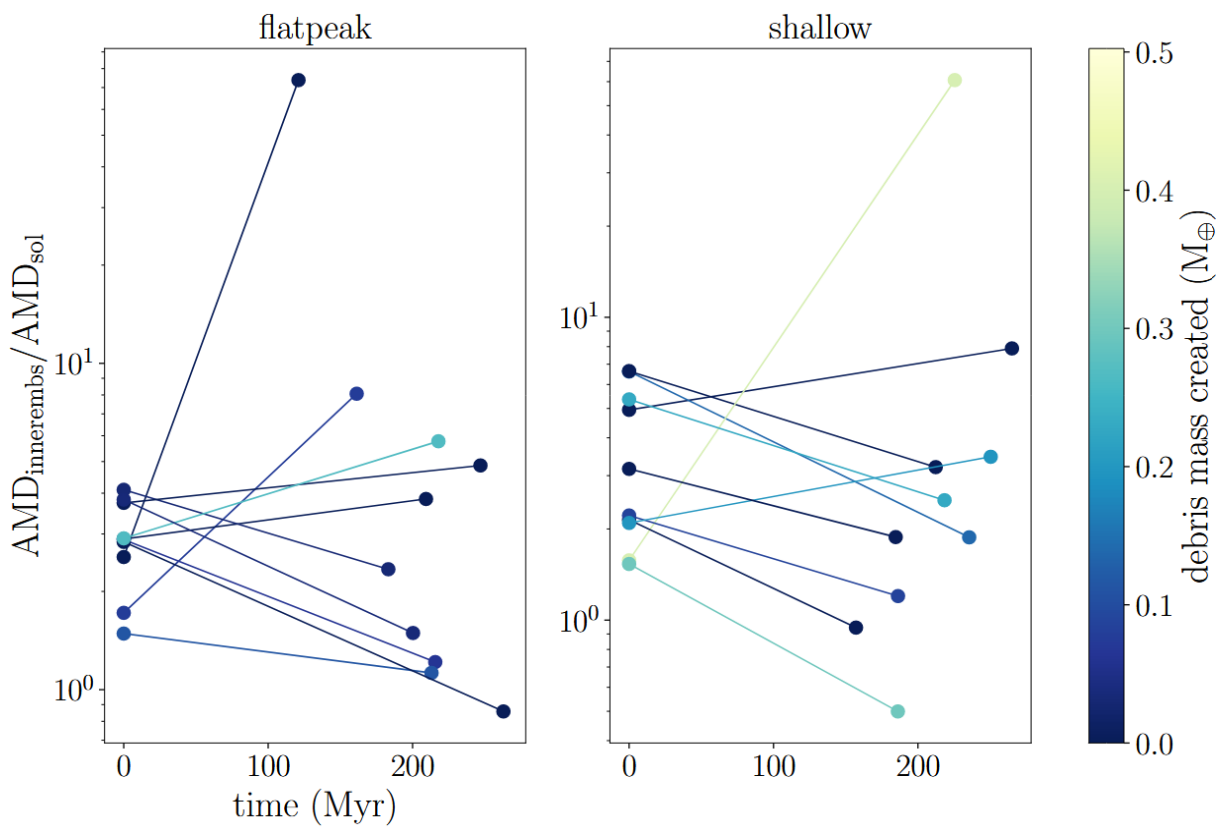

Figure 8 – For the 10 systems obtained from the "flatpeak1" (left) and "inner-shallow" (right) gas-disk profiles, the figure depicts the AMD at the end of the "case 1" giant planet instability (the dot placed at "time 0") and the final AMD after a timespan ranging from 100 to almost 300 My. The initial and final AMDs of each system are connected by a line. The color denotes the total mass of generated debris, according to the scale on the right hand side of the figure.

## 5. Conclusions

We performed 170 *N*-body simulations for at least 210 Myr to study the formation of the terrestrial planets from a ring of planetesimals. The first 10 Myr of the simulations have been performed with the GPU N-body code GENGA (Grimm et al., 2022; Grimm and Stadel,

2014) assuming four different types of gas disks that lead to the convergent migration of planetary embryos towards 1 AU. The disk was removed on an e-folding time of 1 Myr. After the gas disk fully dissipated, we took the results from the GENGA simulations at 10 Myr and redistributed the particles with slight change in their velocity vectors to generate new sets of initial conditions. These have then been put into a modified version of SyMBA, named as iSyMBA (Roig et al., 2021), which enacted the instability of the giant planets previously obtained in different simulations. In this way we could study the effect of the giant planet instability on the formation of the terrestrial planets. More specifically, we considered two different evolutions of giant planets during the instability (case1 and case3 of Nesvorný et al., 2013), both leading to correct final orbits of the giant planets. We enacted these instabilities at three different times (15 Myr, 60 Myr and 100 Myr) to study how sensitive the terrestrial planet formation process is to this parameter.

The results are encouraging. In 50% of the simulations, we form two planets of about the Earth's mass in the Earth-Venus region (0.5 to 1.2 AU) on orbits with small eccentricities and inclinations. The mass-distance distribution reproduces quite well that of the real terrestrial planets, particularly in the disk dubbed "shallow inner", pointing towards small Mercury in region < 0.5 AU and Mars analogues in region 1.4 to 1.65 AU at its extremes. Nevertheless, only about 10% of the simulations form a strict Mars with mass < 0.2 $M_{Earth}$. This success rate is about 20 to 30 % lower than in previous simulations of terrestrial planet formation from a ring of bodies (Hansen, 2009; Nesvorný et al., 2021; Izidoro et al., 2022), but these simulations ignored the growth and spreading of the embryos during the gas disk phase which our simulations show to be an important effect.

We then switched our attention to study how the timing of the giant planet instability affects the growth of Earth-sized planets. We found that 50 % of the last giant impacts on Earth-sized planets occur within 30 Myr after the giant planet instability. Thus, an early giant planet instability (at 15 Myr) is likely to result in a last giant impact occurring before 45 Myr, while chronological constraints indicate that the Moon-forming event occurred between 75 and 150 Myr (Kleine and Nimmo, 2024). A later last giant impact on Earth, in the 75 – 150 My range, would be more likely triggered by a late giant planet instability. Nevertheless, late last giant impacts are possible with a non-negligible probability also in simulations assuming an early giant planet instability. So, our study cannot constrain the timing of the instability.

A late last giant impact, however, is likely to be associated with a deficit in the late veneer mass delivered to Earth, because there are not enough planetesimals left in the system. Similarly, the angular momentum deficit (AMD) of the final terrestrial planets tends to be too high. We find that a minority of simulations are consistent with the late veneer mass and AMD with a giant impact occurring between 75 and 100 My.

We have discussed some possibilities to increase the late veneer mass and damp the AMD and tested the effect of generating debris during giant impacts, which could potentially enhance the dynamical friction on the resulting planetary system. We found that the effects are too small to significantly reduce the AMD of systems affected by a too late giant impact. The late veneer mass on Earth and the AMD of the terrestrial planet system suggest that the Moon-forming event occurred 50-80 My after the beginning of the Solar system (Fig. 7, right panel). The indications of a later lunar formation mostly come from chronometers dating the crystallization of the lunar magma ocean (Kleine and Nimmo, 2024). We propose that the partial remelting of the lunar mantle due to a high-eccentricity episode (Cuk et al., 2016) could explain these late crystallization ages and be compatible with an earlier formation of our satellite as indicated by our study.

**Acknowledgement**

This work was funded by the Deutsche Forschungsgemeinschaft (SFB-TRR 170, subproject B6, project no. 209) and the ERC project N. 101019380 "HolyEarth". This work was granted access to the HPC resources of IDRIS under the allocation 2023-A0120413416 made by GENCI. The debris-generating simulations were performed on the Sunnyvale computer at the Canadian Institute for Theoretical Astro-physics (CITA).

**References**
o	Adachi, I., Hayashi, C., Nakazawa, K., 1976. The Gas Drag Effect on the Elliptic Motion of a Solid Body in the Primordial Solar Nebula. Prog. Theor. Phys. 56, 1756–1771. https://doi.org/10.1143/PTP.56.1756
o	Asphaug, E., Agnor, C.B., Williams, Q., 2006. Hit-and-run planetary collisions. Nature 439, 155–160. https://doi.org/10.1038/nature04311
o	Barboni, M., Boehnke, P., Keller, B., Kohl, I.E., Schoene, B., Young, E.D., McKeegan, K.D., 2017. Early formation of the Moon 4.51 billion years ago. Sci. Adv. 3, e1602365. https://doi.org/10.1126/sciadv.1602365
o	Benz, W., Slattery, W.L., Cameron, A.G.W., 1988. Collisional stripping of Mercury's mantle. Icarus 74, 516–528. https://doi.org/10.1016/0019-1035(88)90118-2
o	Borg, L.E., Brennecka, G.A., Kruijer, T.S., 2022. The origin of volatile elements in the Earth–Moon system. Proc. Natl. Acad. Sci. 119, e2115726119. https://doi.org/10.1073/pnas.2115726119
o	Borg, L.E., Connelly, J.N., Boyet, M., Carlson, R.W., 2011. Chronological evidence that the Moon is either young or did not have a global magma ocean. Nature 477, 70–72. https://doi.org/10.1038/nature10328
o	Borg, L.E., Gaffney, A.M., Kruijer, T.S., Marks, N.A., Sio, C.K., Wimpenny, J., 2019. Isotopic evidence for a young lunar magma ocean. Earth Planet. Sci. Lett. 523, 115706. https://doi.org/10.1016/j.epsl.2019.07.008
o	Brasil, P.I.O., Roig, F., Nesvorny, D., Carruba, V., Aljbaae, S., Huaman, M.E. 2016. Dynamical dispersal of primordial asteroid families. Icarus 266, 142–151. doi:10.1016/j.icarus.2015.11.015
o	Brasser, R., Morbidelli, A., Gomes, R., Tsiganis, K., Levison, H.F., 2009. Constructing the secular architecture of the solar system II: the terrestrial planets. Astron. Astrophys. 507, 1053–1065. https://doi.org/10.1051/0004-6361/200912878
o	Brasser, R., Walsh, K.J., Nesvorný, D., 2013. Constraining the primordial orbits of the terrestrial planets. Mon. Not. R. Astron. Soc. 433, 3417–3427. https://doi.org/10.1093/mnras/stt986
o	Brož, M., Chrenko, O., Nesvorný, D., Dauphas, N., 2021. Early terrestrial planet formation by torque-driven convergent migration of planetary embryos. Nat. Astron. 5, 898–902. https://doi.org/10.1038/s41550-021-01383-3
o	Canup, R.M., 2012. Forming a Moon with an Earth-like Composition via a Giant Impact. Science 338, 1052–1055. https://doi.org/10.1126/science.1226073
o	Canup, R.M., Asphaug, E., 2001. Origin of the Moon in a giant impact near the end of the Earth's formation. Nature 412, 708–712. https://doi.org/10.1038/35089010
o	Carrera, D., Johansen, A., Davies, M.B., 2015. How to form planetesimals from mm-sized chondrules and chondrule aggregates. Astron. Astrophys. 579, A43. https://doi.org/10.1051/0004-6361/201425120
o	Carter, P.J., Leinhardt, Z.M., Elliott, T., Walter, M.J., Stewart, S.T. 2015. Compositional Evolution during Rocky Protoplanet Accretion. The Astrophysical Journal 813.

doi:10.1088/0004-637X/813/1/72
- Carter, P.J., Stewart, S.T. 2022. Did Earth Eat Its Leftovers? Impact Ejecta as a Component of the Late Veneer. The Planetary Science Journal 3. doi:10.3847/PSJ/ac6095
- Chambers, J.E. 2013. Late-stage planetary accretion including hit-and-run collisions and fragmentation. Icarus 224, 43–56. doi:10.1016/j.icarus.2013.02.015
- Chau, A., Reinhardt, C., Helled, R., Stadel, J., 2018. Forming Mercury by Giant Impacts. Astrophys. J. 865, 35. https://doi.org/10.3847/1538-4357/aad8b0
- Chou, C.-L., 1978. Fractionation of Siderophile Elements in the Earth's Upper Mantle. Proc. 9th Lunar Planet. Sci. Conf. Houst. TX 219–230.
- Clement, M.S., Kaib, N.A., Raymond, S.N., Walsh, K.J., 2018. Mars' growth stunted by an early giant planet instability. Icarus 311, 340–356. https://doi.org/10.1016/j.icarus.2018.04.008
- Clement, M.S., Kaib, N.A., Chambers, J.E. 2020. Embryo Formation with GPU Acceleration: Reevaluating the Initial Conditions for Terrestrial Accretion. The Planetary Science Journal 1. doi:10.3847/PSJ/ab91aa
- Clement, M.S., Raymond, S.N., Kaib, N.A., Deienno, R., Chambers, J.E., Izidoro, A. 2021a. Born eccentric: Constraints on Jupiter and Saturn's pre-instability orbits. Icarus 355. doi:10.1016/j.icarus.2020.114122
- Clement, M.S., Deienno, R., Kaib, N.A., Izidoro, A., Raymond, S.N., Chambers, J.E. 2021b. Born extra-eccentric: A broad spectrum of primordial configurations of the gas giants that match their present-day orbits. Icarus 367. doi:10.1016/j.icarus.2021.114556
- Clement, M.S., Chambers, J.E., Kaib, N.A., Raymond, S.N., Jackson, A.P. 2023. Mercury's formation within the early instability scenario. Icarus 394. doi:10.1016/j.icarus.2023.115445
- Connelly, J.N., Nemchin, A.A., Merle, R.E., Snape, J.F., Whitehouse, M.J., Bizzarro, M., 2022. Calibrating volatile loss from the Moon using the U-Pb system. Geochim. Cosmochim. Acta 324, 1–16. https://doi.org/10.1016/j.gca.2022.02.026
- Crespi, S. and 6 colleagues 2021. Protoplanet collisions: Statistical properties of ejecta. Monthly Notices of the Royal Astronomical Society 508, 6013–6022. doi:10.1093/mnras/stab2951
- Crespi, S., Ali-Dib, M., Dobbs-Dixon, I. 2024. Protoplanet collisions: new scaling laws from SPH simulations. arXiv e-prints. doi:10.48550/arXiv.2402.07803
- Delbo, M., Avdellidou, C., Morbidelli, A., 2019. Ancient and primordial collisional families as the main sources of X-type asteroids of the inner main belt. Astron. Astrophys. 624, A69. https://doi.org/10.1051/0004-6361/201834745
- Delbo, M., Walsh, K., Bolin, B., Avdellidou, C., Morbidelli, A., 2017. Identification of a primordial asteroid family constrains the original planetesimal population. Science 357, 1026–1029. https://doi.org/10.1126/science.aam6036
- Deienno, R., Walsh, K.J., Kretke, K.A., Levison, H.F. 2019. Energy Dissipation in Large Collisions - No Change in Planet Formation Outcomes. The Astrophysical Journal 876. doi:10.3847/1538-4357/ab16e1
- Duncan, M.J., Levison, H.F., Lee, M.H., 1998. A Multiple Time Step Symplectic Algorithm for Integrating Close Encounters. Astron. J. 116, 2067. https://doi.org/10.1086/300541
- Elkins-Tanton, L.T., 2008. Linked magma ocean solidification and atmospheric growth for Earth and Mars. Earth Planet. Sci. Lett. 271, 181–191. https://doi.org/10.1016/j.epsl.2008.03.062
- Flock, M., Fromang, S., Turner, N.J., Benisty, M., 2017. 3D Radiation Nonideal Magnetohydrodynamical Simulations of the Inner Rim in Protoplanetary Disks. Astrophys. J. 835, 230. https://doi.org/10.3847/1538-4357/835/2/230
- Gaffney, A.M., Borg, L.E., 2014. A young solidification age for the lunar magma ocean. Geochim. Cosmochim. Acta 140, 227–240. https://doi.org/10.1016/j.gca.2014.05.028
- Genda, H., Kokubo, E., Ida, S. 2012. Merging Criteria for Giant Impacts of Protoplanets. The Astrophysical Journal 744. doi:10.1088/0004-637X/744/2/137
- Genda, H., Brasser, R., Mojzsis, S.J. 2017. The terrestrial late veneer from core

- disruption of a lunar-sized impactor. Earth and Planetary Science Letters 480, 25–32. doi:10.1016/j.epsl.2017.09.041
- Gomes, R., Levison, H.F., Tsiganis, K., Morbidelli, A., 2005. Origin of the cataclysmic Late Heavy Bombardment period of the terrestrial planets. Nature 435, 466–469. https://doi.org/10.1038/nature03676
- Grimm, S.L., Stadel, J.G., 2014. THE GENGA CODE: GRAVITATIONAL ENCOUNTERS INN-BODY SIMULATIONS WITH GPU ACCELERATION. Astrophys. J. 796, 23. https://doi.org/10.1088/0004-637X/796/1/23
- Grimm, S.L., Stadel, J.G., Brasser, R., Meier, M.M.M., Mordasini, C., 2022. GENGA. II. GPU Planetary N-body Simulations with Non-Newtonian Forces and High Number of Particles. Astrophys. J. 932, 124. https://doi.org/10.3847/1538-4357/ac6dd2
- Halliday, A.N., 2008. A young Moon-forming giant impact at 70–110 million years accompanied by late-stage mixing, core formation and degassing of the Earth. Philos. Trans. R. Soc. Math. Phys. Eng. Sci. 366, 4163–4181. https://doi.org/10.1098/rsta.2008.0209
- Hansen, B.M.S., 2009. FORMATION OF THE TERRESTRIAL PLANETS FROM A NARROW ANNULUS. Astrophys. J. 703, 1131–1140. https://doi.org/10.1088/0004-637X/703/1/1131
- Hyodo, R., Genda, H., Brasser, R., 2021. Modification of the composition and density of Mercury from late accretion. Icarus 354, 114064. https://doi.org/10.1016/j.icarus.2020.114064
- Izidoro, A., Raymond, S.N., Pierens, A., Morbidelli, A., Winter, O.C., Nesvorny, D. 2016. The Asteroid Belt as a Relic from a Chaotic Early Solar System. The Astrophysical Journal 833. doi:10.3847/1538-4357/833/1/40
- Izidoro, A., Dasgupta, R., Raymond, S.N., Deienno, R., Bitsch, B., Isella, A., 2022. Planetesimal rings as the cause of the Solar System's planetary architecture. Nat. Astron. 6, 357–366. https://doi.org/10.1038/s41550-021-01557-z
- Jacobson, S.A., Morbidelli, A., 2014. Lunar and terrestrial planet formation in the Grand Tack scenario. Philos. Trans. R. Soc. Math. Phys. Eng. Sci. 372, 20130174. https://doi.org/10.1098/rsta.2013.0174
- Jacobson, S.A., Morbidelli, A., Raymond, S.N., O'Brien, D.P., Walsh, K.J., Rubie, D.C., 2014. Highly siderophile elements in Earth's mantle as a clock for the Moon-forming impact. Nature 508, 84–87. https://doi.org/10.1038/nature13172
- Kaib, N.A., Chambers, J.E., 2016. The fragility of the terrestrial planets during a giant-planet instability. Mon. Not. R. Astron. Soc. 455, 3561–3569. https://doi.org/10.1093/mnras/stv2554
- Kaib, N.A., Cowan, N.B., 2015. The feeding zones of terrestrial planets and insights into Moon formation. Icarus 252, 161–174. https://doi.org/10.1016/j.icarus.2015.01.013
- Kleine, T., Touboul, M., Bourdon, B., Nimmo, F., Mezger, K., Palme, H., Jacobsen, S.B., Yin, Q.-Z., Halliday, A.N., 2009. Hf–W chronology of the accretion and early evolution of asteroids and terrestrial planets. Geochim. Cosmochim. Acta, The Chronology of Meteorites and the Early Solar System 73, 5150–5188. https://doi.org/10.1016/j.gca.2008.11.047
- Kleine, T. & Walker, R. J. Tungsten Isotopes in Planets. *Ann. Rev. Earth Planet. Sci.* 45, 389–417 (2017).
- Kleine, T., Nimmo, F., 2024. Origin of the Earth. Treaties in Geochemistry. Submitted
- Kobayashi, H., Isoya, K., Sato, Y. 2019. Importance of Giant Impact Ejecta for Orbits of Planets Formed during the Giant Impact Era. The Astrophysical Journal 887. doi:10.3847/1538-4357/ab5307
- Kokubo, E., Genda, H. 2010. Formation of Terrestrial Planets from Protoplanets Under a Realistic Accretion Condition. The Astrophysical Journal 714, L21–L25. doi:10.1088/2041-8205/714/1/L21
- Korenaga, J., Marchi, S. 2023. Vestiges of impact-driven three-phase mixing in the chemistry and structure of Earth's mantle. Proceedings of the National Academy of Science 120. doi:10.1073/pnas.2309181120
- Kruijer, T.S., Archer, G.J., Kleine, T., 2021. No 182W evidence for early Moon

formation. Nat. Geosci. 14, 714–715. https://doi.org/10.1038/s41561-021-00820-2

o   Kruijer, T.S., Burkhardt, C., Budde, G., Kleine, T., 2017. Age of Jupiter inferred from the distinct genetics and formation times of meteorites. Proc. Natl. Acad. Sci. 114, 6712–6716. https://doi.org/10.1073/pnas.1704461114

o   Leinhardt, Z.M., Stewart, S.T. 2012. Collisions between Gravity-dominated Bodies. I. Outcome Regimes and Scaling Laws. The Astrophysical Journal 745. doi:10.1088/0004-637X/745/1/79

o   Li, R., Youdin, A.N., 2021. Thresholds for Particle Clumping by the Streaming Instability. Astrophys. J. 919, 107. https://doi.org/10.3847/1538-4357/ac0e9f

o   Liu, B., Raymond, S.N., Jacobson, S.A., 2022. Early Solar System instability triggered by dispersal of the gaseous disk. Nature 604, 643–646. https://doi.org/10.1038/s41586-022-04535-1

o   Lock, S.J., Stewart, S.T., Petaev, M.I., Leinhardt, Z., Mace, M.T., Jacobsen, S.B., Cuk, M., 2018. The Origin of the Moon Within a Terrestrial Synestia. J. Geophys. Res. Planets 123, 910–951. https://doi.org/10.1002/2017JE005333

o   Lykawka, P.S., 2020. Can narrow discs in the inner Solar system explain the four terrestrial planets? Mon. Not. R. Astron. Soc. 496, 3688–3699. https://doi.org/10.1093/mnras/staa1625

o   Lykawka, P.S., Ito, T. 2023. Terrestrial planet and asteroid belt formation by Jupiter-Saturn chaotic excitation. Scientific Reports 13. doi:10.1038/s41598-023-30382-9

o   Mezger, K., Maltese, A., Vollstaedt, H., 2021. Accretion and differentiation of early planetary bodies as recorded in the composition of the silicate Earth. Icarus 365, 114497. https://doi.org/10.1016/j.icarus.2021.114497

o   Mojzsis, S.J., Brasser, R., Kelly, N.M., Abramov, O., Werner, S.C., 2019. Onset of Giant Planet Migration before 4480 Million Years Ago. Astrophys. J. 881, 44. https://doi.org/10.3847/1538-4357/ab2c03

o   Morbidelli, A., Baillié, K., Batygin, K., Charnoz, S., Guillot, T., Rubie, D.C., Kleine, T., 2022. Contemporary formation of early Solar System planetesimals at two distinct radial locations. Nat. Astron. 6, 72–79. https://doi.org/10.1038/s41550-021-01517-7

o   Morbidelli, A., Bottke, W.F., Nesvorný, D., Levison, H.F., 2009. Asteroids were born big. Icarus 204, 558–573. https://doi.org/10.1016/j.icarus.2009.07.011

o   Morbidelli, A., Tsiganis, K., Crida, A., Levison, H.F., Gomes, R., 2007. Dynamics of the Giant Planets of the Solar System in the Gaseous Protoplanetary Disk and Their Relationship to the Current Orbital Architecture. Astron. J. 134, 1790. https://doi.org/10.1086/521705

o   Morbidelli, A., Wood, B.J., 2015. Late Accretion and the Late Veneer, in: The Early Earth. American Geophysical Union (AGU), pp. 71–82. https://doi.org/10.1002/9781118860359.ch4

o   Morishima, R., Schmidt, M.W., Stadel, J., Moore, B., 2008. Formation and Accretion History of Terrestrial Planets from Runaway Growth through to Late Time: Implications for Orbital Eccentricity. Astrophys. J. 685, 1247. https://doi.org/10.1086/590948

o   Morishima, R., Stadel, J., Moore, B. 2010. From planetesimals to terrestrial planets: N-body simulations including the effects of nebular gas and giant planets. Icarus 207, 517–535. doi:10.1016/j.icarus.2009.11.038

o   Nesvorný, D., 2018. Dynamical Evolution of the Early Solar System. Annu. Rev. Astron. Astrophys. 56, 137–174. https://doi.org/10.1146/annurev-astro-081817-052028

o   Nesvorný, D., Morbidelli, A., 2012. STATISTICAL STUDY OF THE EARLY SOLAR SYSTEM'S INSTABILITY WITH FOUR, FIVE, AND SIX GIANT PLANETS. Astron. J. 144, 117. https://doi.org/10.1088/0004-6256/144/4/117

o   Nesvorný, D., Roig, F.V., Deienno, R., 2021. The Role of Early Giant-planet Instability in Terrestrial Planet Formation. Astron. J. 161, 50. https://doi.org/10.3847/1538-3881/abc8ef

o   Nesvorný, D., Roig, F.V., Vokrouhlický, D., Bottke, W.F., Marchi, S., Morbidelli, A., Deienno, R., 2023. Early bombardment of the moon: Connecting the lunar crater record to the terrestrial planet formation. Icarus 399, 115545.

https://doi.org/10.1016/j.icarus.2023.115545

o Nesvorný, D., Vokrouhlický, D., Bottke, W.F., Levison, H.F., 2018. Evidence for very early migration of the Solar System planets from the Patroclus–Menoetius binary Jupiter Trojan. Nat. Astron. 2, 878–882. https://doi.org/10.1038/s41550-018-0564-3

o Nesvorný, D., Vokrouhlický, D., Morbidelli, A., 2013. CAPTURE OF TROJANS BY JUMPING JUPITER. Astrophys. J. 768, 45. https://doi.org/10.1088/0004-637X/768/1/45

o O'Brien, D.P., Morbidelli, A., Levison, H.F., 2006. Terrestrial planet formation with strong dynamical friction. Icarus 184, 39–58. https://doi.org/10.1016/j.icarus.2006.04.005

o Ogihara, M., Kokubo, E., Suzuki, T.K., Morbidelli, A., 2018. Formation of the terrestrial planets in the solar system around 1 au via radial concentration of planetesimals. Astron. Astrophys. 612, L5. https://doi.org/10.1051/0004-6361/201832654

o Roig, F., Nesvorný, D., Deienno, R., Garcia, M.J., 2021. isymba: a symplectic massive bodies integrator with planets interpolation. Mon. Not. R. Astron. Soc. 508, 4858–4868. https://doi.org/10.1093/mnras/stab2874

o Roig, F., Nesvorný, D., DeSouza, S.R., 2016. JUMPING JUPITER CAN EXPLAIN MERCURY'S ORBIT. Astrophys. J. Lett. 820, L30. https://doi.org/10.3847/2041-8205/820/2/L30

o Rubie, D. C., Nathan, G., Dale, K., Nakajima, M., Jennings, E.S., Golabek, G., Jacobson, S. Morbidelli, A. 2024. Effects of metal-silicate fractionation mechanisms on $^{182}$W isotopic evolution during Earth's core formation, PNAS, submitted

o Rudge, J.F., Kleine, T., Bourdon, B., 2010. Broad bounds on Earth's accretion and core formation constrained by geochemical models. Nat. Geosci. 3, 439–443. https://doi.org/10.1038/ngeo872

o Scora, J., Valencia, D., Morbidelli, A., Jacobson, S. 2020. Chemical diversity of super-Earths as a consequence of formation. Monthly Notices of the Royal Astronomical Society 493, 4910–4924. doi:10.1093/mnras/staa568

o Scora, J., Valencia, D., Morbidelli, A., Jacobson, S. 2022. Rocky Histories: The Effect of High Excitations on the Formation of Rocky Planets. The Astrophysical Journal 940. doi:10.3847/1538-4357/ac9cda

o Shankman, C., Gladman, B.J., Kaib, N., Kavelaars, J.J., Petit, J.M., 2013. A POSSIBLE DIVOT IN THE SIZE DISTRIBUTION OF THE KUIPER BELT'S SCATTERING OBJECTS. Astrophys. J. 764, L2. https://doi.org/10.1088/2041-8205/764/1/L2

o Stewart, S.T., Leinhardt, Z.M. 2012. Collisions between Gravity-dominated Bodies. II. The Diversity of Impact Outcomes during the End Stage of Planet Formation. The Astrophysical Journal 751. doi:10.1088/0004-637X/751/1/32

o Suzuki, T.K., Ogihara, M., Morbidelli, A., Crida, A., Guillot, T., 2016. Evolution of protoplanetary discs with magnetically driven disc winds. Astron. Astrophys. 596, A74. https://doi.org/10.1051/0004-6361/201628955

o Svetsov, V., 2011. Cratering erosion of planetary embryos. Icarus 214, 316–326. https://doi.org/10.1016/j.icarus.2011.04.026

o Tanaka, H., Takeuchi, T., Ward, W.R., 2002. Three-Dimensional Interaction between a Planet and an Isothermal Gaseous Disk. I. Corotation and Lindblad Torques and Planet Migration. Astrophys. J. 565, 1257. https://doi.org/10.1086/324713

o Tanaka, H., Ward, W.R., 2004. Three-dimensional Interaction between a Planet and an Isothermal Gaseous Disk. II. Eccentricity Waves and Bending Waves. Astrophys. J. 602, 388. https://doi.org/10.1086/380992

o Thiemens, M.M., Sprung, P., Fonseca, R.O.C., Leitzke, F.P., Münker, C., 2019. Early Moon formation inferred from hafnium–tungsten systematics. Nat. Geosci. 12, 696–700. https://doi.org/10.1038/s41561-019-0398-3

o Tsiganis, K., Gomes, R., Morbidelli, A., Levison, H.F., 2005. Origin of the orbital architecture of the giant planets of the Solar System. Nature 435, 459–461. https://doi.org/10.1038/nature03539

o Ueda, T., Ogihara, M., Kokubo, E., Okuzumi, S., 2021. Early Initiation of Inner Solar System Formation at the Dead-zone Inner Edge. Astrophys. J. Lett. 921, L5. https://doi.org/10.3847/2041-8213/ac2f3b

- Walker, R.J., 2009. Highly siderophile elements in the Earth, Moon and Mars: Update and implications for planetary accretion and differentiation. Geochemistry 69, 101–125. https://doi.org/10.1016/j.chemer.2008.10.001
- Walsh, K.J., Levison, H.F., 2016. TERRESTRIAL PLANET FORMATION FROM AN ANNULUS. Astron. J. 152, 68. https://doi.org/10.3847/0004-6256/152/3/68
- Walsh, K.J., Morbidelli, A., Raymond, S.N., O'Brien, D.P., Mandell, A.M., 2011. A low mass for Mars from Jupiter's early gas-driven migration. Nature 475, 206–209. https://doi.org/10.1038/nature10201
- Walsh, K.J., Levison, H.F. 2019. Planetesimals to terrestrial planets: Collisional evolution amidst a dissipating gas disk. Icarus 329, 88–100. doi:10.1016/j.icarus.2019.03.031
- Wetherill, G.W., Stewart, G.R., 1989. Accumulation of a swarm of small planetesimals. Icarus 77, 330–357. https://doi.org/10.1016/0019-1035(89)90093-6
- Woo, J.M.Y., Brasser, R., Grimm, S.L., Timpe, M.L., Stadel, J., 2022. The terrestrial planet formation paradox inferred from high-resolution N-body simulations. Icarus 371, 114692. https://doi.org/10.1016/j.icarus.2021.114692
- Woo, J.M.Y., Morbidelli, A., Grimm, S.L., Stadel, J., Brasser, R., 2023. Terrestrial planet formation from a ring. Icarus 396, 115497. https://doi.org/10.1016/j.icarus.2023.115497
- Yang, C.-C., Johansen, A., Carrera, D., 2017. Concentrating small particles in protoplanetary disks through the streaming instability. Astron. Astrophys. 606, A80. https://doi.org/10.1051/0004-6361/201630106
- Youdin, A.N., Goodman, J., 2005. Streaming Instabilities in Protoplanetary Disks. Astrophys. J. 620, 459. https://doi.org/10.1086/426895